\documentstyle[12pt,epsf]{article}
\input epsf.tex
 1

 1

 1

\begin{document}

\noindent
\begin{titlepage}
\vspace{12pt}
\begin{center}
\begin{large}

             {\bf Relativistic Kinetic Vertex in Positronium \\}

\end{large}
\vspace{22pt} {\bf  S.M. Zebarjad $^{a,c}$ \footnote{E-mail :
zebarjad@physics.susc.ac.ir} and\,\,} {\bf M. Haghighat$^{b,c}$
\footnote{E-mail : mansour@cc.iut.ac.ir}}

\vspace{12pt} {\it
\vspace{8pt}
$^a$ Physics Department,  Shiraz University, Shiraz 71454,  Iran, \\
$^b$ Physics Department,  Isfahan University of Technology (IUT), Isfahan 84154,  Iran, \\
$^c$Institute for Studies in Theoretical Physics and Mathematics\\
             (IPM), Tehran 19395, Iran.\\

\vspace{0.3cm}}
\end{center}

\vspace{2cm} \abstract{We derive Spinless Salpeter  equation for
the positronium using the NRQED Lagrangian.  Consequently, we
consider the Spinless Salpeter wavefunction instead of the
Schrodinger wavefunction to show that the NRQED calculation can be
done easier. We also discuss that
 the singularity of the Spinless Salpeter wavefunction at the origin is necessary
to cancel the ultraviolet divergence in the NRQED calculations. }
\end{titlepage}
\null

\section{Introduction}
A modern method to calculate a non-relativistic bound state
problem is based on the Effective Field Theory (NRQED) which was
proposed by Caswell and Lapage\cite{caswell}. This is the
advantage of this method that provides a set of systematic rules
(power-counting) that allows  an easy identification of all terms
that contribute to a certain order in the bound state
calculations.

In NRQED calculation, relativistic kinetic vertex needs more
attention respect to the other interactions\cite{hfs6pra}. In this
paper, we consider this vertex in the unperturbed part of the
Hamiltonian as well as the Coulomb interaction to obtain the
positronium energy correction. For this purpose we derive the
Spinless Salpeter (SS) equation for the positronium with Coulomb
potential using the NRQED Lagrangian in Sec. 2. The SS
eigenfunctions and eigenvalues in Sec. 3 and 4 are used to
describe the contribution of the relativistic vertex correction to
the energy spectrum of the positronium at the order of $\alpha^4$
and $\alpha^6$.

\section{Spinless Salpeter Equation}
There are many equations incorporating relativistic kinematics
which all have the same non-relativistic limit\cite{lich}.
However, if an expansion is made in powers of the momentum, the
various equations differ in terms next to leading order.
Furthermore, most of them are phenomenological equations and
which are not derivable from field theory without making gross
approximations. An approach which seems promising is the Spinless
Salpeter equation to obtain spectrum of a two-body system.  This
equation, with appropriate assumptions, can be derived from the
Bethe-Salpeter equation\cite{hen} which is manifestly covariant
and can be obtained directly from quantum field theory.  We now
consider the Spinless Salpeter Equation\cite{sal} as follows:
\begin{equation}
\label{SS11}\bigg\{ \sqrt{p_1^2+m_1^2}+\sqrt{p_2^2+m_2^2} \bigg\}
\Psi=(E-V)\Psi,
\end{equation}
where for positronium, in the center of mass frame, one has
\begin{equation}
\label{SS12} 2\sqrt{p^2+m^2} \Psi=(E-V)\Psi,
\end{equation}
or
\begin{equation}
 \label{SS}\bigg\{
{\nabla}^2+\frac{(E-V({\bf r}))^2}{4}-m^2 \bigg\} \Psi( {\bf
{r})=0 },
\end{equation}
where $m$ is the mass of the particle and $V({\bf {r})}$ is the
time component of a 4-vector potential where in the Coulomb case $
V({\bf r})=\frac{-\alpha}{r}$. In the non-relativistic limit,
Eq.(\ref{SS}) leads to a Schrodinger equation with the effective
potential $V$. In fact this equation gives all relativistic
kinetic corrections to the Schrodinger equation with such
effective potential. However we can derive Eq.(\ref{SS}) using
NRQED Lagrangian. The momentum space equation of motion for an
off-shell, time independent $e^+ e^- e^+e^-$ four point function
in the center-of-mass frame is:
\begin{equation}
\bigg[\,
 \frac{\mbox{\boldmath $p$}^2}{m_e}
\,-\,E \,\bigg]\, \tilde G(\mbox{\boldmath $p$},\mbox{\boldmath
$q$};s) \,+\, \int\frac{d^3 \mbox{\boldmath
$q$}^\prime}{(2\,\pi)^3}\, \tilde V(\mbox{\boldmath
$p$},\mbox{\boldmath $q$}^\prime)\, \tilde G(\mbox{\boldmath
$q$},\mbox{\boldmath $q$}^\prime;s) \, = \,
(2\,\pi)^3\,\delta^{(3)}(\mbox{\boldmath $p$}-\mbox{\boldmath
$q$}) \,, 
\end{equation}
where
\begin{equation}
E \, \equiv \, \sqrt{s}-2\,m_e,
\end{equation}
is the center-of-mass energy relative to the electron--positron
threshold and  $\tilde V(\mbox{\boldmath $p$},\mbox{\boldmath
$q$})$, the potential, is defined as
\begin{eqnarray}
\tilde V(\mbox{\boldmath $p$},\mbox{\boldmath $q$}) & = & \tilde
V_{\mbox{\tiny Coul}}(\mbox{\boldmath $p$},\mbox{\boldmath $q$}) +
\tilde V_{\mbox{\tiny BF}}(\mbox{\boldmath $p$},
  \mbox{\boldmath $q$}) +
\tilde V_{4}(\mbox{\boldmath $p$},
  \mbox{\boldmath $q$}) +
\tilde V_{4\,\mbox{\tiny der}}(\mbox{\boldmath $p$},
  \mbox{\boldmath $q$}) +
\delta\tilde H_{\mbox{\tiny kin}}(\mbox{\boldmath $p$},
\mbox{\boldmath $q$}) \,, 
\end{eqnarray}
 introduced in Ref\cite{hfs6pra}.
  Considering $ \delta\tilde
H_{\mbox{\tiny kin}}(\mbox{\boldmath $p$})$  to all orders will
lead to the following equation:
\begin{equation} \bigg[\,
 \frac{\mbox{\boldmath $p$}^2}{m_e}+\frac{\mbox{\boldmath $p$}^4}{4m_e^3}
\,+\, ...-\,E \,\bigg]\, \tilde G(\mbox{\boldmath
$p$},\mbox{\boldmath $q$};s)+(V_{\mbox{\tiny
Coul}}+V^\prime)G(\mbox{\boldmath $p$},\mbox{\boldmath $q$};s) \,
= \, (2\,\pi)^3\,\delta^{(3)}(\mbox{\boldmath $p$}-\mbox{\boldmath
$q$}) \,, 
\end{equation}
or
\begin{equation}
\bigg[\,2\sqrt{p^2+m_e^2}-\sqrt{s}
 \,\bigg]\, \tilde G(\mbox{\boldmath $p$},\mbox{\boldmath
$q$};s)+(V_{\mbox{\tiny Coul}}+V^\prime)G(\mbox{\boldmath
$p$},\mbox{\boldmath $q$};s) = \,
(2\,\pi)^3\,\delta^{(3)}(\mbox{\boldmath $p$}-\mbox{\boldmath
$q$}) \,, \label{NNLOSchroedinger}
\end{equation}
where
\begin{equation}
(V_{\mbox{\tiny Coul}}+V^\prime)G(\mbox{\boldmath
$p$},\mbox{\boldmath $q$};s)\,=\,\int\frac{d^3 \mbox{\boldmath
$q$}^\prime}{(2\,\pi)^3}\, \tilde V_{\mbox{\tiny
Coul}}(\mbox{\boldmath $p$},\mbox{\boldmath $q$}^\prime)\, \tilde
G(\mbox{\boldmath $q$},\mbox{\boldmath $q$}^\prime;s) \,+\,
\int\frac{d^3 \mbox{\boldmath $q$}^\prime}{(2\,\pi)^3}\, \tilde
V^\prime(\mbox{\boldmath $p$},\mbox{\boldmath $q$}^\prime)\,
\tilde G(\mbox{\boldmath $q$},\mbox{\boldmath $q$}^\prime;s) \,,
\end{equation}
 and
\begin{eqnarray}
\tilde V^\prime(\mbox{\boldmath $p$},\mbox{\boldmath $q$}) & = &
 \tilde V_{\mbox{\tiny BF}}(\mbox{\boldmath $p$},
  \mbox{\boldmath $q$}) +
\tilde V_{4}(\mbox{\boldmath $p$},
  \mbox{\boldmath $q$}) +
\tilde V_{4\,\mbox{\tiny der}}(\mbox{\boldmath $p$},
  \mbox{\boldmath $q$})
 \,. \label{momentumspacepotentials}
\end{eqnarray}
Now equation (\ref{NNLOSchroedinger}) can be rewritten as,
\begin{equation}
\bigg[\,2\sqrt{p^2+m_e^2}-(\sqrt{s}-V_{\mbox{\tiny Coul}})
 \,\bigg]\, \tilde G(\mbox{\boldmath $p$},\mbox{\boldmath
$q$};s)+V^\prime\,G(\mbox{\boldmath $p$},\mbox{\boldmath $q$};s) =
\, (2\,\pi)^3\,\delta^{(3)}(\mbox{\boldmath $p$}-\mbox{\boldmath
$q$}) \,. \label{NNLOSchroedinger1}
\end{equation}
Since the Coulomb potential  and the relativistic kinetic vertices
are the only  interactions that must be treated exactly, we can
ignore $V^\prime$ in the above equation which  leads to the
Eq.(\ref{SS}) in momentum space. Therefore one can start with the
wavefunction of the SS equation with Coulomb potential and
determines the corrections by perturbation as is usual in NRQED.
   Whereas the effect of relativistic kinetic vertices  is in the SS
wavefunction of Eq.(\ref{SS}).  To this end by substitution
$\Psi=\frac{U}{r}$ in the radial part of Eq.(\ref{SS}) one
obtains:
\begin{equation}
\label{SS1}\bigg\{{\frac{d^2}{{d\rho^2}}}+ {\frac{2}{\rho}}-\frac{%
(\ell+\kappa)(\ell+\kappa+1)}{\rho^2} -\frac{1}{(n+\kappa)^2}\bigg\}%
U(\rho)=0,
\end{equation}
where $$\kappa=-(\ell+1/2)+\sqrt{(\ell+1/2)^2-\alpha^2/4}\,\,\,\,
\&\,\,\,\, \rho=E\alpha r/4.$$ The eigenvalues and eigenfunctions
of Eq.(\ref{SS1}) are respectively\cite{fac}\cite{grei}:
\begin{eqnarray}
E_{n,\ell}=2m\bigg[1+\frac{\alpha^2}{4(n+\kappa)^2}\bigg]^{-1/2},  \label{energy} \\
U_{n+\kappa}^{\kappa}=N\bigg(\frac{2}{n+\kappa}\bigg)^{\frac{1}{2}+\kappa+n}
\Gamma^{-\frac{1}{3}}(2n+2\kappa+1)\,\,\,\rho^{n+\kappa}\exp{(\frac{-\rho}{n+\kappa}}),
\end{eqnarray}
in which $N$ is an appropriate normalization constant.  We can now
expand $\frac{\Psi_{SS}}{\Psi_{Sch}}$ in terms of $\alpha$ (when
$n=1$, $\ell=0$) as
\begin{equation}
\Psi_{SS}({\bf r})=\Psi_{Sch}({\bf r})\bigg\{1+\alpha^2\,\bigg(
-\frac{1}{4}\gamma+\frac{5}{8}-\frac{1}{4}\ln{(mr \alpha)}\bigg)+\dots%
\bigg\}, \label{GKG}
\end{equation}
where $\gamma$ is the Euler number and $\Psi_{Sch}({\bf r})$ is
the ground state wavefunction for the Schrodinger equation with
Coulomb potential,
\begin{equation}
\Psi_{Sch}({\bf r})=\frac{1}{\sqrt{\pi}} \bigg(\frac{m\alpha}{2}
\bigg)^{3/2}\exp{(-m \alpha r /2)}. \label{schr}
\end{equation}
 The Fourier transformation of Eq.(\ref{GKG}) is
\begin{eqnarray}
\psi_{SS}({\bf p})&=&\psi({\bf p})_{Sch}+\alpha^2  \varphi({\bf
p})+\dots, 
\end{eqnarray}
where $ \psi_{Sch}({\bf p})$ and $\varphi ({\bf p})$ are
\begin{eqnarray}
\psi_{Sch}({\bf p})&=&\frac{8\pi^{1/2}(m\alpha/2)^{5/2}}{[p^2+(m\alpha/2)^2]^2},\\
\varphi({\bf p})&=&\nonumber -\frac{(m\alpha/2)^{3/2}
\sqrt{\pi}}{((m\alpha/2)^{2}+p^2)^2p}\bigg\{\arctan(\frac{2p}{m\alpha})
\bigg(p^2-(m\alpha/2)^2\bigg) \nonumber \\
&      &
+p(m\alpha/2)\ln\bigg(\frac{(m\alpha/2)^{2}+p^2}{m^2\alpha^2}\bigg)
+3p(m\alpha/2)\bigg\}.
\end{eqnarray}
We are now ready to calculate positronium energy corrections at the order of $%
\alpha^4$ and  $\alpha^6$, using SS wavefunction, Eq.(\ref{GKG}).
\section{$\alpha^4$ Positronium Energy Corrections}
The known vertices in NRQED Lagrangian  are given in \cite{lamb}.
The power-counting of NRQED\cite{patrickPC} specifies all the
diagrams contributing to the energy shift at the order of
$\alpha^4$. These diagrams are all shown in Fig.(\ref{a4}) which
can be calculated \cite{mythesis} to
give the well-known positronium energy correction at the order of $\alpha^4$%
\cite{Itzykson}.
  Contribution of relativistic kinetic corrections at this order are shown in
Figs.(\ref{a4}(i,j)) which reads:
\begin{figure}[h]
\centerline{\epsfxsize=6in\epsffile{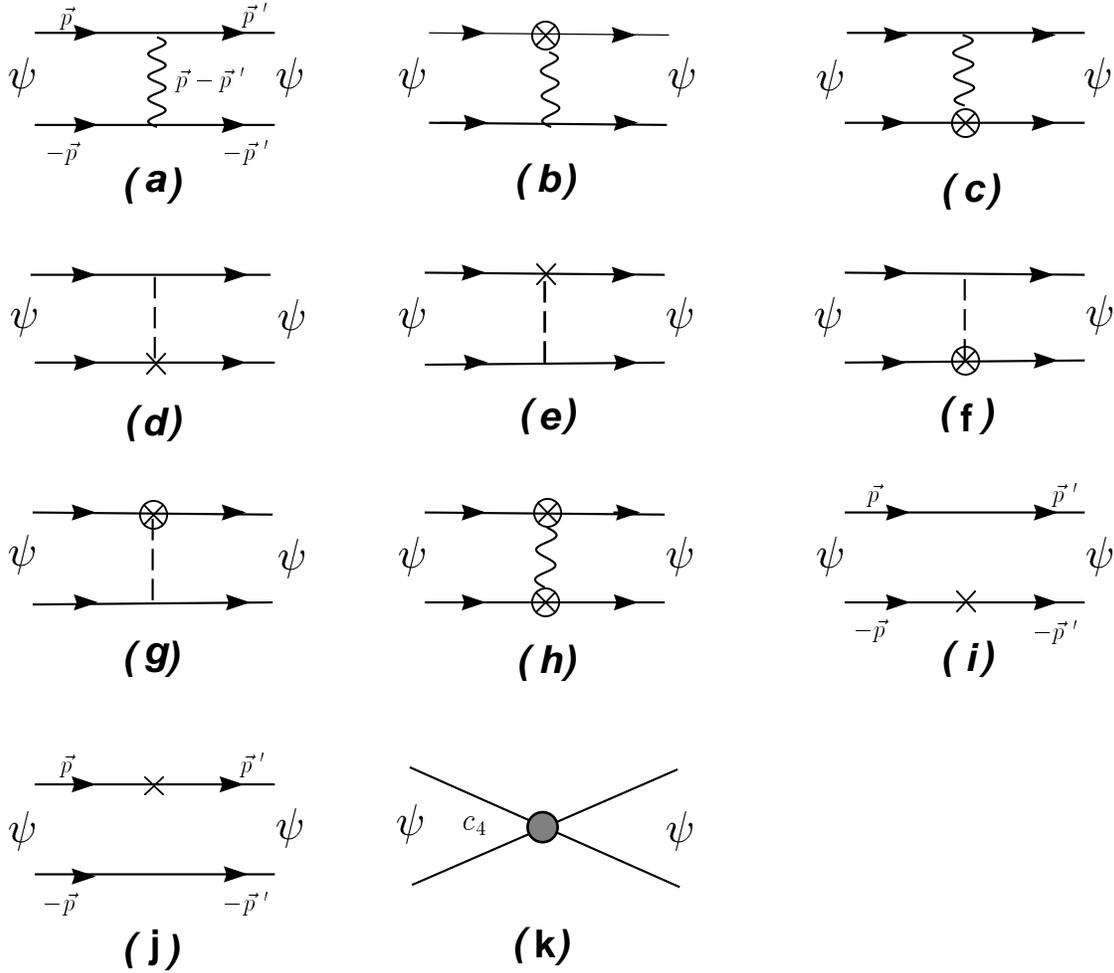}} \caption{ The
whole bound state diagrams contributing to positronium energy
corrections at the order of  $\alpha^4$. $\psi$ is schrodinger
wavefunction and $c_4$ is the leading order result coming from
tree-level matching $(c_4=c^{(0)}_4=-\frac{\pi\alpha}{m^2})$.}
\label{a4}
\end{figure}
\begin{eqnarray}
\Delta E_{i}+\Delta
E_{j}&=&2\int\frac{d^3pd^3p'}{(2\pi)^6}\psi^*({\bf p'}) \bigg[
\frac{-p^4(2\pi)^3  \delta({\bf p}-{\bf
p'})}{8m^3}\bigg]\psi(\bf{p}) \nonumber \\
&=&\frac{-m\alpha^4}{16n^3(\ell+1/2)}+\frac{3m \alpha^4}{64n^4}.
\label{ra4}
\end{eqnarray}
To obtain positronium energy correction at the order of $\alpha^4$
using SS equation, we should omit Figs.(\ref{a4}(i,j)) and
calculate the remaining diagrams in Fig.(\ref{a4}) using SS
wavefunction. That is obviously leads to the previous result at
the order of $\alpha^4$ and also extra pieces which start at
higher order. These terms are irrelevant to the order of our
interest. Meanwhile the result of Eq.(\ref{ra4}) can be easily
obtained by expanding the Eq.(\ref{energy}) in terms of  $\alpha$,
\begin{equation}
E_{n,\ell}=2m\bigg[1+\frac{\alpha^2}{4(n+\kappa)^2}\bigg]^{-1/2}=2m
-\frac{m\alpha^2}{4n^2}-\frac{m\alpha^4}{16n^3(\ell+1/2)}+\frac{3m
\alpha^4}{64n^4}+\dots.
 \label{enrexpan}
\end{equation}
\section{$\alpha^6$ Positronium Energy Correction}
The full calculation of the positronium energy correction at the
order of $\alpha^6$, using NRQED has been done in reference
\cite{fullcal}. In this paper, we focus on the positronium
Hyperfine Splitting (HFS) at the order of $\alpha^6$ coming from
one-photon annihilation \cite{hfs6}\cite{hfs6pra}. To be more
specific  in the way that the SS wavefunction  can be useful  in
the bound state calculation, we first briefly review the NRQED
method in the following subsection.
\subsection{Matching  and bound state energy shift at NNLO  }
\begin{figure}
\centerline{\epsfxsize=6in\epsffile{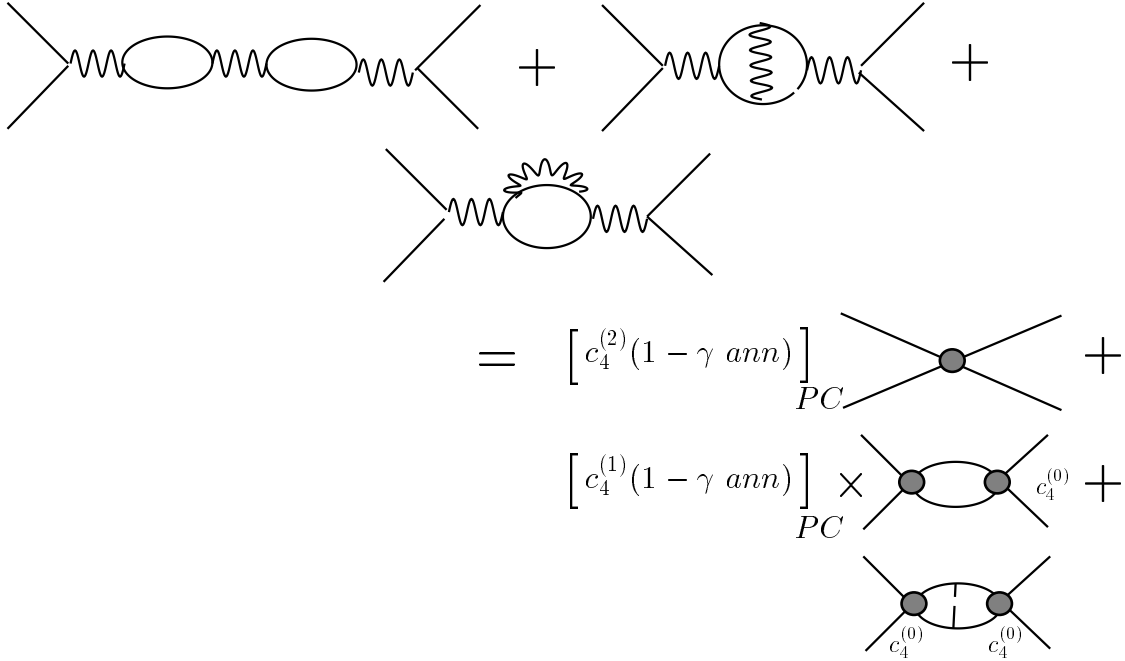}} \caption{The
matching which determines $c^{(2)}_4(1-\gamma\,\, ann)_{PC}$ while
$c^{(1)}_4(1-\gamma\,\, ann)_{PC}$ can be similarly obtained
from  the one-loop matching procedure.} \label{match1}
\end{figure}

\begin{figure}
\centerline{\epsfxsize=6in\epsffile{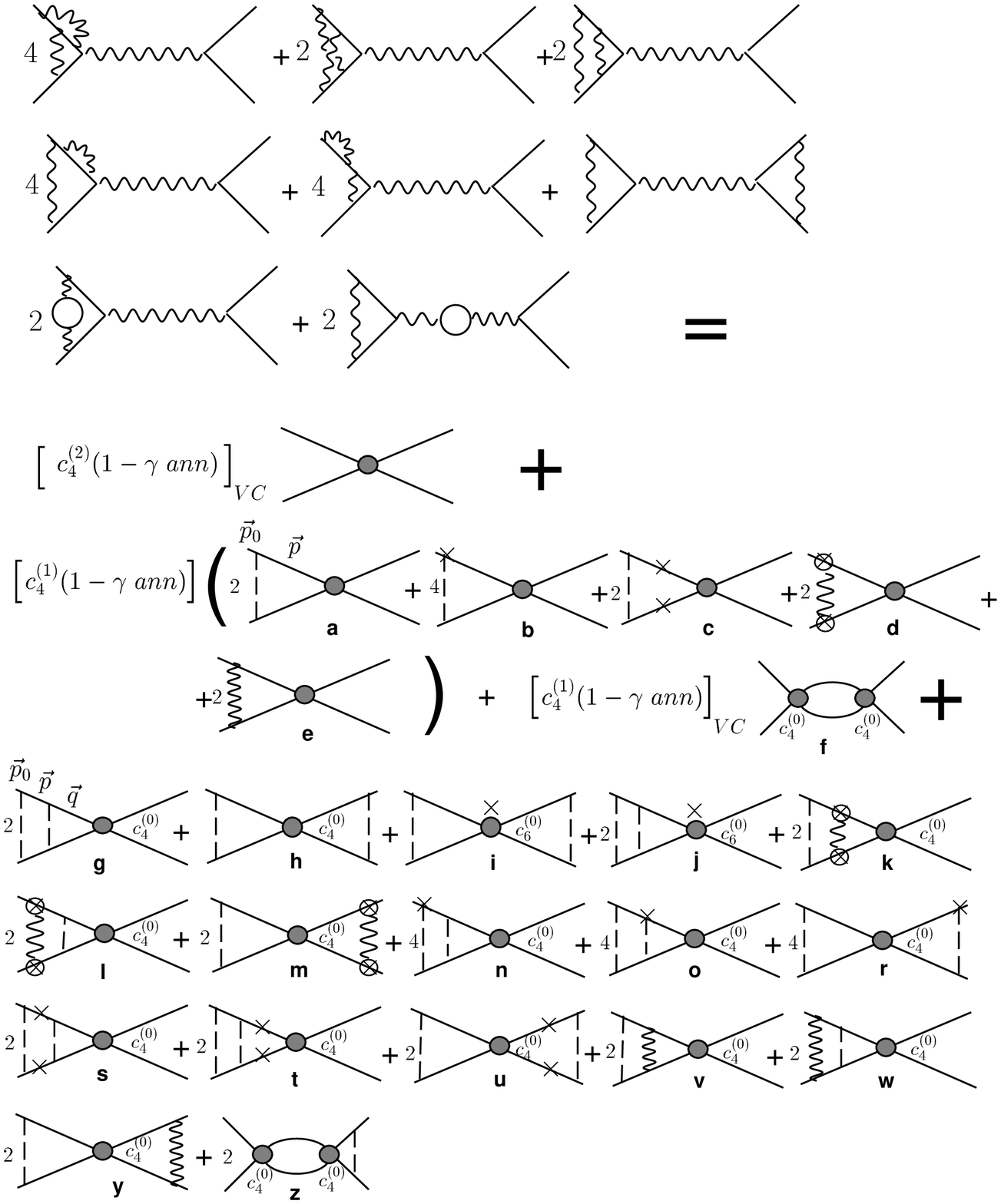}} \caption{The
matching which determines $c^{(2)}_4(1-\gamma\,\, ann)_{VC}$ while
$c^{(1)}_4(1-\gamma\,\, ann)_{VC}$ can be similarly obtained
from  the one-loop matching procedure.
 } \label{match2}
\end{figure}

\begin{figure}[h]
\centerline{\epsfxsize=6in\epsffile{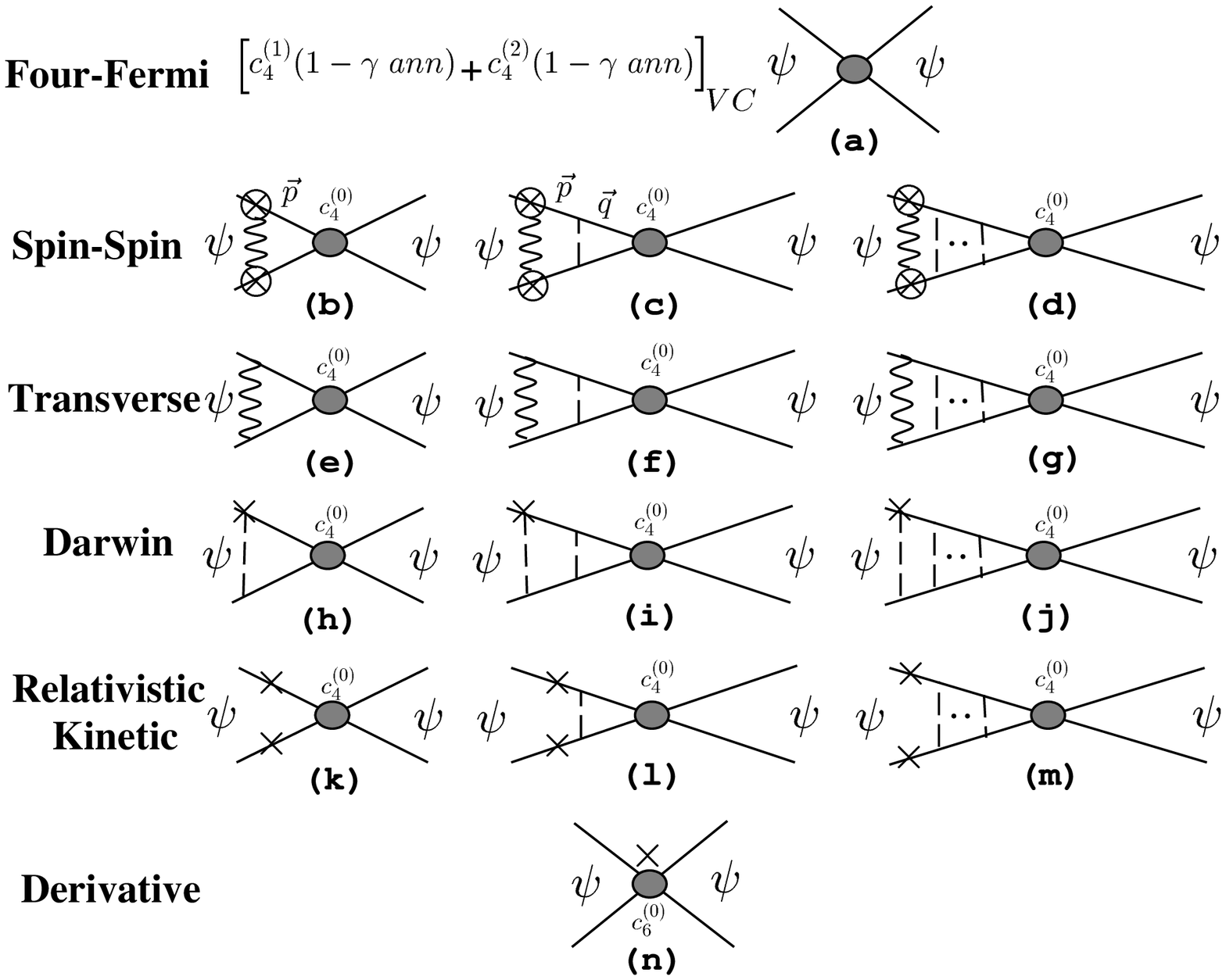}} \caption{All
the bound state diagrams which contribute to $\alpha^6$ except the
diagrams with the Double Annihilation Interaction. $c_6^{(0)}$ is
called the Derivative interaction  comes from the Taylor expansion
of each vertex of the one photon annihilation diagram as well as
the photon propagator correction which is equal to
$\frac{-2\pi\alpha}{3m^4}$. } \label{a6a}
\end{figure}

\begin{figure}[h]
\centerline{\epsfxsize=6in\epsffile{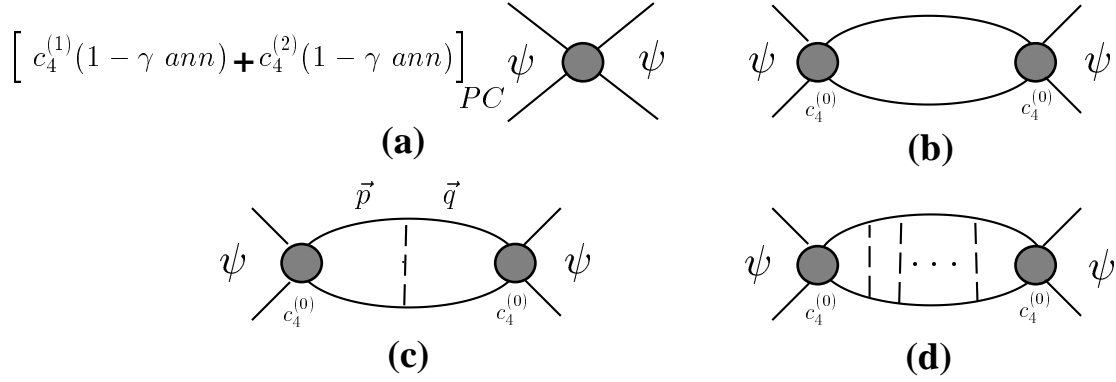}} \caption{The
whole Double Annihilation bound state diagrams.} \label{a6b}
\end{figure}

 Since the single photon  annihilation of electron-positron
occurs  in $S=1$ state, one should calculate the diagrams which
contain spin-1 Four-Fermion Vertex, $c_4(1-\gamma\,\, ann)$. We
can write $ c_4(1-\gamma\,\, ann)=c^{(0)}_4(1-\gamma\,\,
ann)+c^{(1)}_4(1-\gamma\,\, ann)+\dots$ where the superscript
"(0)" and "(1)"  indicate that the coefficients $ c_4(1-\gamma\,\,
ann)$  have been derived using  tree-level and one-loop matching,
respectively. To obtain the contribution of HFS at the order
$\alpha^6$, we should perform matching at two-loop level to get
$c^{(2)}_4(1-\gamma\,\, ann)$. For this purpose, it is more
convenient to write:
\begin{equation}
c^{(2)}_4(1-\gamma\,\, ann)=c^{(2)}_4(1-\gamma\,\,
ann)_{PC}+c^{(2)}_4(1-\gamma\,\, ann)_{VC},
\label{matching}
\end{equation}
where PC stands for contribution coming from propagator
corrections while VC for vertex corrections. Each terms in right
hand-side of Eq.(\ref{matching}) can be obtained from
Figs.(\ref{match1}-\ref{match2}).
 All the NRQED diagrams which contribute to HFS at the order of
$\alpha^6$ can be identified using the NRQED power-counting
rules\cite{patrickPC}. These are shown in
Figs.(\ref{a6a}-\ref{a6b}) which are completely calculated in
\cite{mythesis}\cite{hfs6pra}. Diagrams of Figs.(\ref{a6a}(k,l,m))
which contain relativistic vertex corrections result
in\cite{mythesis}:
\begin{eqnarray}
Fig(\ref{a6a},k)+Fig(\ref{a6a},l)+Fig(\ref{a6a},m)=\frac{m\alpha
^6}8\ln \bigg( \frac \Lambda {m\alpha }\bigg)+\frac 1{32}m\alpha
^6+\frac {m\alpha^5}{ 4\pi} \frac{\Lambda} m.\label{5}
\end{eqnarray}
\begin{figure}[h]
\centerline{\epsfxsize=5in\epsffile{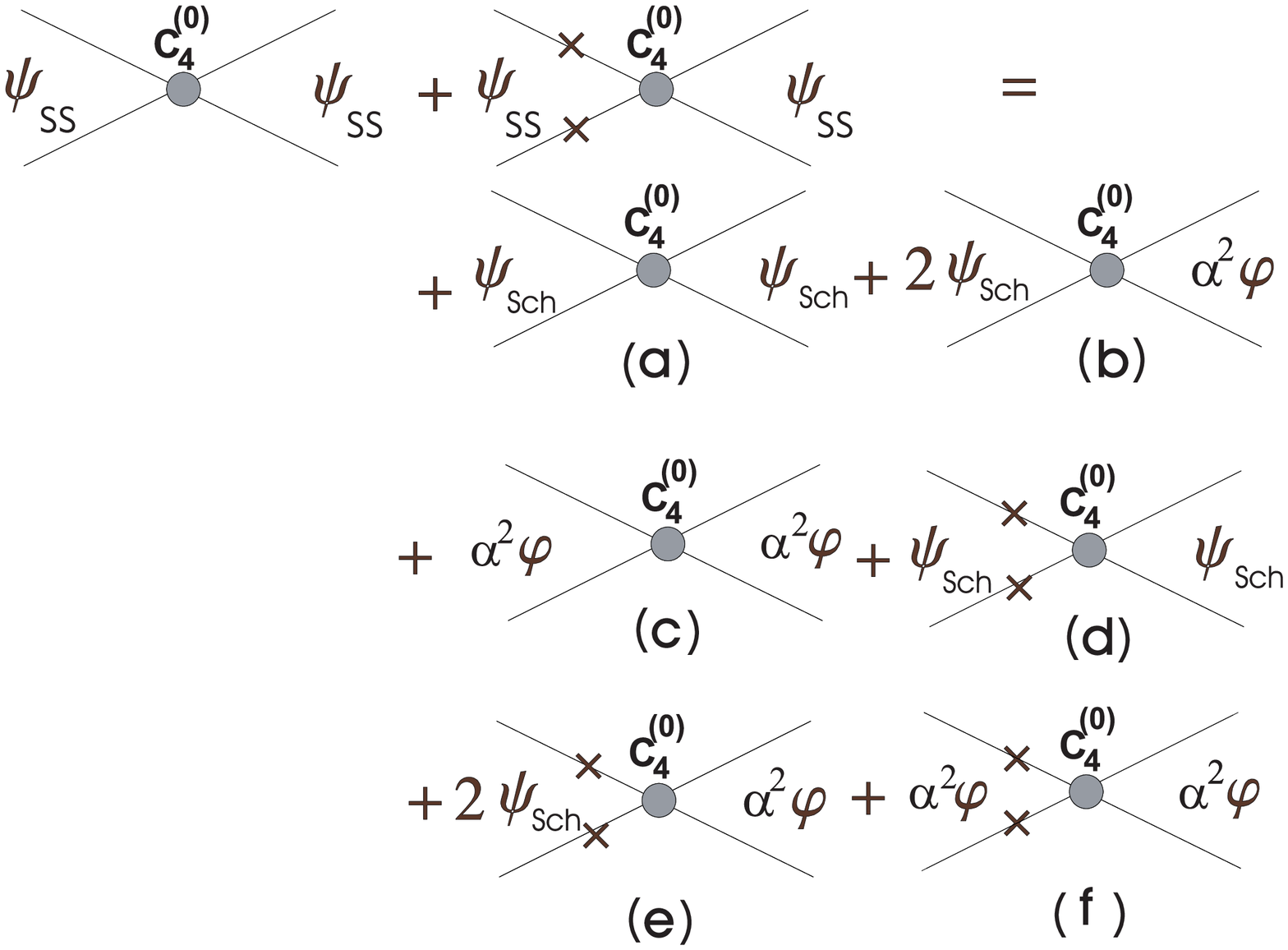}} \caption{The
diagrams at the order $\alpha^6$ and higher coming from one
photon annihilation at the order $\alpha^4$.} \label{KG1}
\end{figure}
\subsection{NNLO bound state energy shift using SS wavefunction}
As  shown in section 2,  the whole relativistic kinetic vertex as
well as Coulomb vertex can be considered in unperturbed part of
the Hamiltonian, Eq.(\ref{NNLOSchroedinger1}). Therefore, one
should expect to omit relativistic kinetic vertex  and using SS
wavefunction  instead of Schrodinger wavefunction,  in all order
of perturbation, to obtain the same results  within the framework
of NRQED. That is basically reduce the number of counter terms
involve in this calculation \cite{smirnov}.
 Since our goal is to obtain the final result using SS
wavefunction, we should omit Figs.(\ref{a6a}(k,l,m)) and
calculate the remaining diagrams in Fig(\ref{a6a}-\ref{a6b}) with
the SS wavefunction. Straightforward calculations shows that we
have the previous result for these diagrams and also extra pieces
at higher order than $\alpha^6$. At first glance, it seems that
there is no way to get the value of Figs.(\ref{a6a}k,l,m) which
we have omitted, but if we consider the diagrams which
contributed to HFS at the order of $\alpha^4$ we get some pieces
at the higher order. That is basically due to the fact that we
should replace $\psi_{Sch}$ with $\psi_{SS}$ in Fig.(\ref{a4},k)
and also using the $p/m$ expansion of relativistic propagator.
This means that the Fig.(\ref{a4},k) should be replaced by  the
left hand side of Fig.(\ref{KG1}). It is easy to show that
Fig.(\ref{KG1}(a)) leads to the previous result at the order
$\alpha^4$, while the Figs.(\ref{KG1}(c,e,f)) contributes to the
higher order than $\alpha^6$. The only remaining diagrams
relevant to our calculation are the Figs.(\ref{KG1}(b,d)):
\begin{eqnarray}
Fig(6(b))&=&\frac{4\pi \alpha }{m^2}\int
\frac{d^3p^{\prime}}{(2\pi )^3} \psi_{Sch} ({\bf p^{\prime }})\int
\frac{d^3p}{(2\pi )^3}\alpha ^2{\varphi }( {\bf p})=\frac{m\alpha
^6}{8}\ln \bigg(\frac {\Lambda} {m\alpha }\bigg)+\frac
{5}{32}m\alpha ^6, \label{1}
\end{eqnarray}
\begin{eqnarray}
Fig(6(d))&=&-\frac 1{8}m\alpha ^6+\frac {m\alpha^5}{ 4\pi}
\frac{\Lambda} m. \label{2}
\end{eqnarray}
The sum of Fig.(\ref{KG1}(b)) and Fig.(\ref{KG1}(d)) is just equal
to (\ref{5}). It should be noted here that the divergence in the
first term of (\ref{1}) is a direct consequence of the singularity
of the SS wavefunction (\ref{GKG}).  In fact in the Schrodinger
case Figs.(2(l,m)) are responsible to cancel the logarithmic
divergence coming from the other diagrams while in the SS case
this is due to the SS wavefunction.

\section*{Summary}
In this paper, we have shown that the spinless Salpeter equation
 (\ref{SS}) correctly predicts the energy spectrum of a
non-relativistic two body system such as positronium. In this way
we have ignored the relativistic vertex correction in the bound
state and therefore the calculations are made easier. On the
other hand, although the SS wavefunction is singular at the
origin (see Eq.(\ref{GKG})) but this is a crucial need to cancel
UV divergence in the bound state system, Eq.(\ref{1}).
\section*{Acknowledgement}
The authors gratefully acknowledge the financial support of Shiraz
University research council and  IPM.
\newpage


\sloppy

\raggedright

\def\app#1#2#3{{\it Act. Phys. Pol. }{\bf B #1} (#2) #3}

\def\apa#1#2#3{{\it Act. Phys. Austr.}{\bf #1} (#2) #3}

\def\lhc{Proc. LHC Workshop, CERN 90-10}

\def\npb#1#2#3{{\it Nucl. Phys. }{\bf B #1} (#2) #3}

\def\nP#1#2#3{{\it Nucl. Phys. }{\bf #1} (#2) #3}

\def\plb#1#2#3{{\it Phys. Lett. }{ \bf #1 B } (#2) #3}

\def\prd#1#2#3{{\it Phys. Rev. }{\bf D #1} (#2) #3}

\def\pra#1#2#3{{\it Phys. Rev. }{\bf A #1} (#2) #3}

\def\pR#1#2#3{{\it Phys. Rev. }{\bf #1} (#2) #3}

\def\prl#1#2#3{{\it Phys. Rev. Lett. }{\bf #1} (#2) #3}

\def\prc#1#2#3{{\it Phys. Reports }{\bf #1} (#2) #3}

\def\cpc#1#2#3{{\it Comp. Phys. Commun. }{\bf #1} (#2) #3}

\def\nim#1#2#3{{\it Nucl. Inst. Meth. }{\bf #1} (#2) #3}

\def\pr#1#2#3{{\it Phys. Reports }{\bf #1} (#2) #3}

\def\sovnp#1#2#3{{\it Sov. J. Nucl. Phys. }{\bf #1} (#2) #3}

\def\sovpJ#1#2#3{{\it Sov. Phys. JETP Lett. }{\bf #1} (#2) #3}

\def\jl#1#2#3{{\it JETP Lett. }{\bf #1} (#2) #3}

\def\jet#1#2#3{{\it JETP }{\bf #1} (#2) #3}

\def\zpc#1#2#3{{\it Z. Phys. }{\bf C #1} (#2) #3}

\def\ptp#1#2#3{{\it Prog.~Theor.~Phys.~}{\bf #1} (#2) #3}

\def\nca#1#2#3{{\it Nuovo~Cim.~}{\bf #1A} (#2) #3}

\def\ap#1#2#3{{\it Ann. Phys. }{\bf #1} (#2) #3}

\def\hpa#1#2#3{{\it Helv. Phys. Acta }{\bf #1} (#2) #3}

\def\ijmpa#1#2#3{{\it Int. J. Mod. Phys. }{\bf A #1} (#2) #3}

\def\ZETF#1#2#3{{\it Zh. Eksp. Teor. Fiz. }{\bf #1} (#2) #3}

\def\jmp#1#2#3{{\it J. Math. Phys. }{\bf #1} (#2) #3}

\def\yf#1#2#3{{\it Yad. Fiz. }{\bf #1} (#2) #3}

\def\cjp#1#2#3{{\it Can. J. Phys. }{\bf #1} (#2) #3}


\end{document}